# Digitized Adjoint Method for Inverse Design of Digital Nanophotonic Devices

Xinshu Ren,[†] Weijie Chang,[†] Longhui Lu,[†] Max Yan,[‡] Deming Liu,[†] and Minming Zhang,[*,†]

[†] School of Optical and Electronic Information, Huazhong University of Science and Technology, Wuhan, Hubei 430074, China
[‡] School of Engineering Sciences, KTH Royal Institute of Technology, Stockholm 16440, Sweden

**ABSTRACT:** We present a digitized adjoint method for realizing efficient inverse design of "digital" subwavelength nanophotonic devices. We design a single-mode 3-dB power divider and a dual-mode demultiplexer to demonstrate the digitized adjoint method for single-object and dual-object optimizations, respectively. The optimization comprises three stages, a first stage of continuous variation for an "analog" pattern, a second stage of forced permittivity biasing for a "quasi-digital" pattern, and a third stage for a multi-level digital pattern. Compared with conventional brute-force method, the proposed digitized adjoint method can improve the design efficiency by about 5 times, and the performance optimization can reach approximately the same level using the ternary pattern. The digitized adjoint method takes the advantages of adjoint sensitivity analysis and digital subwavelength structure and creates a new way for efficient and high-performance design of compact digital subwavelength nanophotonic devices. This method could overcome the efficiency bottleneck of the brute-force method that is restricted by the number of pixels of a digital pattern and improve the device performance by extending a conventional binary pattern to a multi-level one, which may be attractive for inverse design of large-scale digital nanophotonic devices.

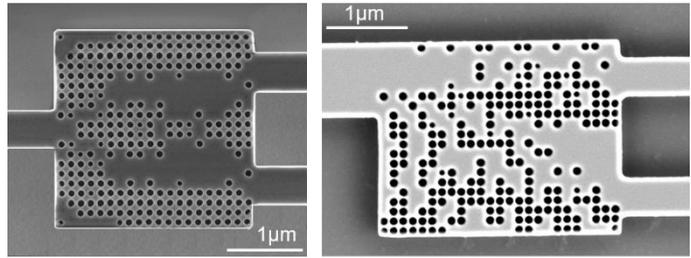

**KEYWORDS:** *silicon photonics, optical devices, inverse design, adjoint method*

A reduction in the size of integrated all-dielectric silicon photonic devices while maintaining a high level of performance is a key challenge for applications with limited physical space such as on-chip optical interconnects. Inverse design approach is recently emerging as a promising way to realize ultracompact and high-performance nanophotonic devices for high-density integration, including nanostructured photonic crystals,[1] wavelength demultiplexers,[2,3] power dividers,[4–7] polarization beamsplitter,[8] polarization rotator,[9] mode demultiplexers,[10,11] mode converters,[12,13] waveguide bend,[14] and twisted light emitter.[15] The basic idea of inverse design is that the design area of a subwavelength photonic device is firstly discretized into different numbers of nanoscale elements, then we use optimization methods to find an optimized refractive index distribution of each element to fulfill the design requirements.

Generally, inverse-designed subwavelength nanophotonic devices may be classified into two categories: "analog" and "digital" nanophotonic devices. Because the dimension of unit element of the analog devices is much smaller than that of the digital ones, the etching patterns of analog devices usually have "arbitrarily" curved boundaries,[1–4,7,10,12] and those of digital ones are normally rectangular- or circular-like shapes.[5,6,8,9,11,13–15] Analog nanophotonic devices offer more degrees of freedom for inverse design at the expenses of higher computational and likely fabrication costs, whereas digital nanophotonic devices have simpler design procedure, easier-to-fabricate patterns, and comparably high performance in various applications. Topology optimization, level-set method, and other gradient-based methods are commonly used for inverse design of analog devices, in which the adjoint method is indispensable to reduce the ultrafine-element-induced tremendous computational cost to a reasonable degree and make the analog inverse design feasible because it could provide the topology or shape gradient information using only a forward and an adjoint (backward) simulations regardless of the number of design elements.[1–4,7,10,12] However, the conventional adjoint method can be hardly applied to inverse design of digital devices because one cannot calculate the gradient of a digital pattern. Simple brute-force methods, such as direct-binary search (DBS) algorithm, are employed mostly and successfully for optimization of digital patterns.[5,6,8,11,13,14] Unfortunately, the number of fully-vectorial 3D simulations in brute-force methods will increase exponentially with the pixel number in a pattern, which may drastically limit the inverse design capability of digital nanophotonic devices.

In this work, we propose a digitized adjoint method for efficient inverse design of digital subwavelength nanophotonic devices. For demonstration purpose, the PhC-like subwavelength structure is used as the base nanostructure of digital nanophotonic devices, and its unit element is a silicon cuboid with a central cylinder filled with silicon or air.[5] The inverse design process of the proposed digitized adjoint method can be divided into three stages. The first stage is geometry-fixed topology optimization. We tune the relative permittivities of all cylinders (i.e., inverse design domain) with a fixed shape continuously and individually, and obtain an optimized analog pattern with "gray" cylinders using adjoint sensitivity analysis. In the second stage, we employ a linear-biasing approach to convert the analog pattern in the first stage to a "quasi-digital" one in which the relative permittivities of most cylinders are close to the two boundary values. The optimization process of this stage is the same as the first stage except that a forced biasing is used to update the cylinders' relative permittivities. In the last stage, we introduce a fabrication-constraint brute-force quantization method to transform the quasi-digital pattern into an $N$-ary digital pattern, in which intermediate cylinders with different "gray" relative permittivities in the quasi-digital pattern are replaced with air cylinders with 45-nm radius, silicon cylinders or air cylinders with $N-2$ different radii on the basis of effective medium theory to try to minimize the performance degradation due to the digitalization process. Here, we use a ternary pattern ($N=3$) based on a 3-level threshold for demonstration.

First, we design a single-mode 3-dB power divider to demonstrate the digitized adjoint method for inverse design under single-objective optimization. In order to compare with the brute-force method, we chose the same device design parameters as ref 5. Specifically, the 3-dB power divider has a compact footprint of 2.6 μm × 2.6 μm and is discretized into 20 × 20 pixels for inverse design, as shown in Figure 1. The device is designed on the 220 nm thick top silicon layer of the silicon-on-insulator (SOI) platform. Each pixel of the PhC-like subwavelength structure is in the shape of a silicon cuboid (130 nm × 130 nm × 220 nm) with a central cylinder with an initial radius of 45 nm and a depth of 140 nm. The depth of air holes in this device is the same as that used in ref. 5. The device layout is axisymmetric. The width of input and output waveguides is 500 nm and the gap between the two output ones is 1 μm wide. The relative permittivities of air and silicon are set to be 1 and 12 in simulations, respectively.

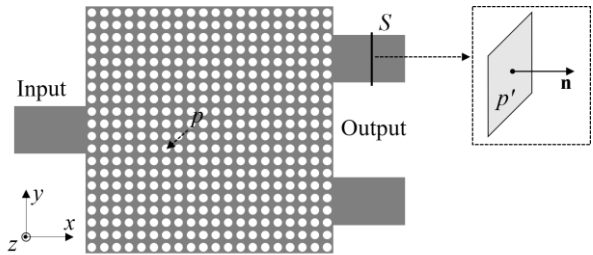

Figure 1. Schematic diagram of the single-mode 3-dB power divider (before optimization).

We define the figure-of-merit (FOM) of the device for inverse design as the transmission into the fundamental transverse electric mode (TE$_0$) mode in the two output waveguides, and the transverse magnetic (TM) mode is neglected for simplify. Because we keep all intermediate patterns axisymmetric, the FOM can be expressed as

$$\text{FOM} = \frac{1}{4} \frac{\left| \int_S \left[ \mathbf{E}(p') \times \overline{\mathbf{H}_0(p')} + \overline{\mathbf{E}_0(p')} \times \mathbf{H}(p') \right] \cdot d\mathbf{S} \right|^2}{\int_S \text{Re}\left[ \mathbf{E}_0(p') \times \overline{\mathbf{H}_0(p')} \right] \cdot d\mathbf{S}} \quad (1)$$

where $S$ is the cross section of the upper output waveguide, $p'$ represents an arbitrary point in $S$, $\mathbf{E}_0$ and $\mathbf{H}_0$ represent the basis electric and magnetic fields of the TE$_0$ mode, and $\mathbf{E}$, $\mathbf{H}$ denote the actual electric and magnetic fields at $S$, respectively. The overline means complex conjugation.

For a cylinder at position $p$, a small change of its relative permittivity, $\delta\varepsilon_r(p)$, introduces an electric dipole moment which leads to a variation of electromagnetic field at $p'$. When we change the relative permittivities of all cylinders simultaneously, the total change of electric field at $p'$ will be a superposition of variations caused by all cylinders. Based on the adjoint method for inverse design of analog patterns,[1,4] the variation in FOM is given by

$$\delta\text{FOM} = 2\varepsilon_0 V \int_\chi \delta\varepsilon_r(p) \text{Re}\left[ \mathbf{E}^A(p) \cdot \mathbf{E}^{old}(p) \right] d^3p \quad (2)$$

where $\chi$ is the design region in all 20 × 20 cylinders with the same and fixed shape, $\varepsilon_0$ is the permittivity of vacuum, $V$ is the volume of a single cylinder, $\mathbf{E}^{old}(p)$ means the electric field at position $p$ before permittivity change, and $\mathbf{E}^A(p)$ represents the adjoint field at $p$. Thus, the path to a gradient-based optimization could be reached by updating relative permittivity of each cylinder in iterations as

$$\delta\varepsilon_r(p) = \text{Re}\left[ \mathbf{E}^A(p) \cdot \mathbf{E}^{old}(p) \right] \quad (3)$$

to ensure that $\delta$FOM maintains positive and thus device performance can be continuously optimized during the iteration process.

In the first stage of digitized adjoint method, all cylinders are uniformly filled with an intermediate material with a relative permittivity of 6.5 and the other area is filled with silicon in the initial pattern. In each iteration, $\mathbf{E}^{old}(p)$ and $\mathbf{E}^A(p)$ can be computed from a forward and an adjoint 3D finite-difference time domain (FDTD) simulations, respectively. Then we calculate $\delta\varepsilon_r(p)$ for each cylinder based on eq 3 and update the relative permittivity of each cylinder through $\varepsilon_r^{new}(p) = \varepsilon_r^{old}(p) + \Delta \cdot \delta\varepsilon_r(p)$, where $\Delta$ is a variable to control the speed of convergence. The relative permittivity change of each cylinder in each iteration should be small enough to realize reliable adjoint sensitivity analysis based on eq 2 derived in the context of perturbation theory. Here, we choose $\Delta = 1/\max\{\delta\varepsilon_r(p)\}$, and convergence is obtained after 50 iterations. The generated analog pattern is shown in Figure 2(a), in which the relative permittivity of each cylinder is distributed between 1 and 12.

In the second stage, we convert the analog pattern in the first stage to a quasi-digital one in which the relative permittivities of most cylinders are close to 1 or 12. We use the same adjoint method to calculate the forward field $\mathbf{E}^{old}(p)$ and adjoint field $\mathbf{E}^A(p)$, but update the relative permittivities with a forced biasing in each iteration, expressed as

$$\varepsilon_r^{biased}(p) = (1+m) \cdot \left[ \varepsilon_r^{new}(p) - 6.5 \right] + 6.5 \quad (4)$$

Here, we set the variable $m$ to be 0.05 to slightly enlarge the distance between the relative permittivity and the central value 6.5, which may force the relative permittivity smaller than 1 or larger than 12, it will be clipped to be 1 or 12, respectively. This linear-biasing approach is an analogy similar to discrete optimization in the inverse design of analog nanophotonic devices,[3] inverse design region which may help to ease the performance degradation caused by discretization of relative permittivities. We calculate the mean square error of the permittivity distribution ($\sigma$) in each iteration as

$$\sigma = \frac{1}{M}\sum_{n=1}^{M} \rho_n, \quad \rho_n = \begin{cases} |\varepsilon_r(n)-1|^2, & 1 \leq \varepsilon_r(n) < 6.5 \\ |\varepsilon_r(n)-12|^2, & 6.5 \leq \varepsilon_r(n) \leq 12 \end{cases} \quad (5)$$

where $M$ is the number of cylinders, $\varepsilon_r(n)$ is the relative permittivity of the $n$th cylinder. The mean square error decreases significantly from 6.42 to 1.01 after 50 iterations, accompanied by a 0.002 dB increase of excess loss at 1550 nm. Figure 2(b) shows the optimized quasi-digital pattern in the second stage.

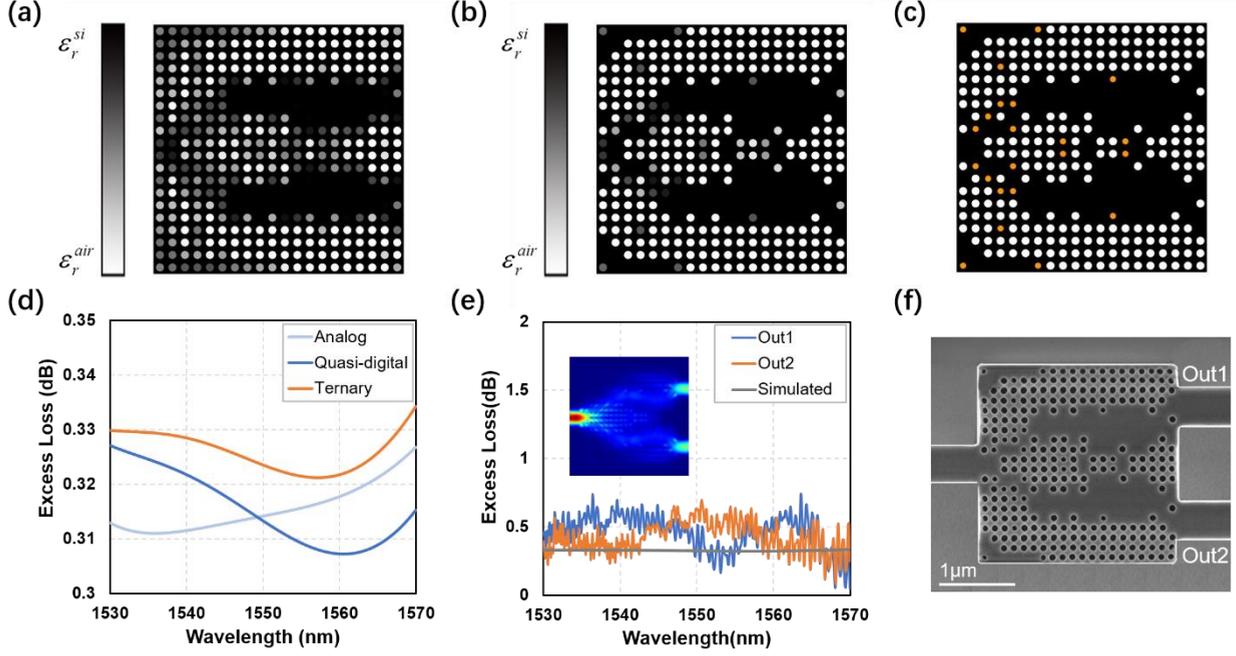

Figure 2. 3-dB power divider. The optimized (a) analog and (b) quasi-digital patterns in the first and second stages, respectively. (c) The optimized ternary pattern in which the smaller air cylinders with a radius of 35 nm is highlighted in orange. (d) Simulated excess loss profiles for the three patterns. (e) Measured excess loss profiles and (f) the SEM image of the fabricated device based on the ternary pattern. The inset in (e) shows the simulated steady-state intensity distribution.

In the third stage, the quasi-digital pattern is transformed to a $N$-ary digital pattern on the basis of effective medium theory. The basic idea is that cylinders with intermediate permittivities and an initial 45-nm radius in the quasi-digital pattern will be replaced with air cylinders with 45-nm radius, air cylinders with an appropriate radius smaller than 45 nm, or silicon cylinders. We use a ternary pattern ($N=3$) based on a 3-level threshold for demonstration. The range of relative permittivity is divided into three segments with two intermediate values of 3.75 and 9.25. Cylinders with relative permittivities larger than 9.25 or smaller than 3.75 in the quasi-digital pattern will be simply filled with silicon or air pattern, respectively. Meanwhile, cylinders with relative permittivities between 3.75 and 9.25 are replaced with smaller air cylinders based on a simple brute-force method. We decrease the radius of all smaller air cylinders from 44 nm to 30 nm with a step of 1 nm, then choose the value corresponding to the best FOM based on 15 rounds of 3D FDTD simulation results. Considering the fabrication constraint, the lower boundary value of the possible radius range is set to be 30 nm. Here, the optimized smaller radius is 35 nm. In our simulation, the feature-size-dependent lag effect of RIE etching depth is always considered and the etch depth of cylinder with a radius of 35 nm is 124 nm.[5] Figure 2(c) shows the optimized ternary pattern in which the smaller air cylinders with the radius of 35 nm are highlighted with the orange color. Notably, if smaller quantization errors of intermediate relative permittivities are preferred, we can use the same method to obtain an $N$-ary digital pattern with $N-2$ intermediate radii (smaller than the initial radius) for small air cylinders based on a $N$-level threshold.

The simulated excess loss profiles of the analog, quasi-digital and ternary patterns are given in Figure 2(d). The average excess loss over 40 nm bandwidth (1530–1570 nm) is 0.32 dB for the analog pattern, and it increases slightly to 0.33 dB for the ternary digital pattern. The measured excess loss profiles and the scanning electron microscope (SEM) picture of the fabricated 3-dB power divider with the optimized ternary pattern are illustrated in Figures 2(e) and 2(f), respectively. The measured average excess loss is 0.44 dB with a fluctuation up to 0.40 dB. The unbalance of excess loss between the two output waveguides is 0.36 dB at most and 0.14 dB in average.

We also design a dual-mode demultiplexer to demonstrate the digitized adjoint method for dual-objective inverse design problem. As shown in Figure 3, the layout of the dual-mode demultiplexer is chosen as same as that in ref 11 for comparison. Specifically, the device has a compact footprint of 2.4 μm × 3 μm. The width of the input and output waveguides are respectively 900 nm and 450 nm. The gap between the two output waveguides is 1.05 μm wide. The design region is discretized into 20 × 25 pixels. Each pixel is a cuboid (120 nm × 120 nm × 220 nm) with a central cylinder. The cross-section radius of each cylinder is initially 45 nm and the depth is set to 220 nm (fully etched).

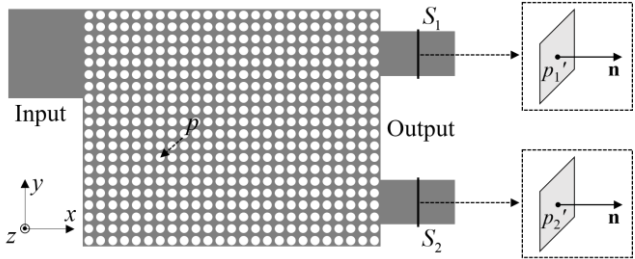

Figure 3. Schematic diagram of the dual-mode demultiplexer (before optimization).

Here, two FOMs are used for inverse design. One (FOM$_1$) represents the transmission into the TE$_0$ mode in the upper output waveguide when TE$_0$ mode is input. The other (FOM$_2$) represents the transmission into the TE$_0$ mode in the lower output waveguide when TE$_1$ mode is input.

In the first stage of the digitized adjoint method, each iteration comprises four simulations for two FOMs. The forward and adjoint simulations for each FOM are similar to those in the optimization of 3-dB power divider, and we firstly obtain the $\delta\varepsilon_{r1}(p)$ and $\delta\varepsilon_{r2}(p)$ based on the independent gradient-based optimizations of two FOMs, respectively. Then we update the relative permittivity of each cylinder as

$$\varepsilon_r^{new}(p)=\varepsilon_r^{old}(p)+\Delta\cdot\frac{1}{2}\left[\delta\varepsilon_{r1}(p)+\delta\varepsilon_{r2}(p)\right] \qquad (6)$$

Here, we set $\Delta = 0.8/\max\{\delta\varepsilon_r(p)\}$, and the convergence of the analog pattern optimization is reached after 100 iterations.

In the second and the last stages, we adopt the same methods used in inverse design of 3-dB power divider to digitize the analog pattern. Specifically, we obtain the quasi-digital pattern of the dual-mode demultiplexer with $\sigma = 0.54$ after 100 iterations in the linear-biasing adjoint optimization based on eq 4. For the optimized ternary pattern, the radius of the small air cylinders is 36 nm.

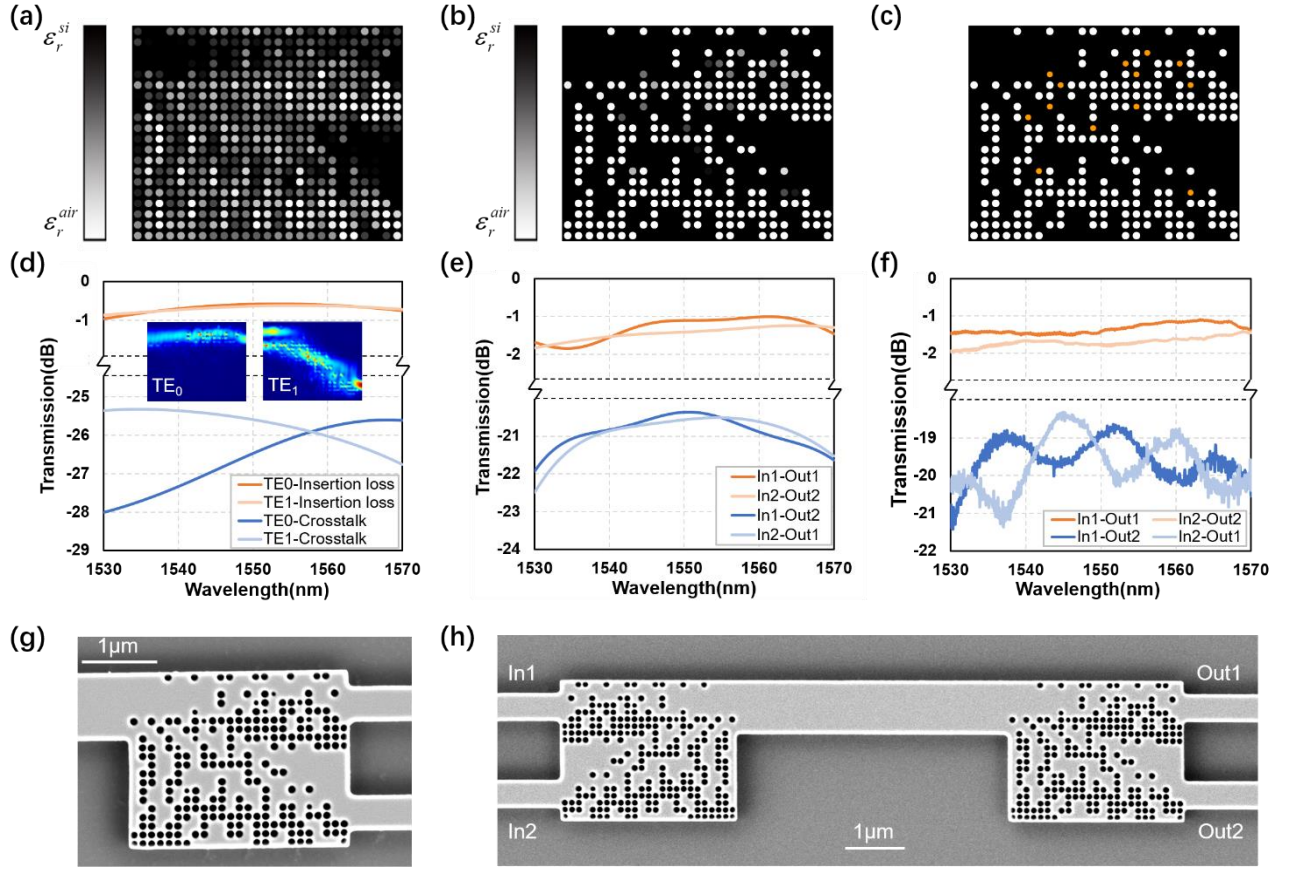

Figure 4. Dual-mode demultiplexer. The optimized (a) analog and (b) quasi-digital patterns in the first and second stages, respectively. (c) The optimized ternary pattern in which the smaller air cylinders with a radius of 36 nm is highlighted in orange. (d) Simulated insertion loss and crosstalk profiles for the ternary pattern. The insets show the simulated steady-state intensity distributions of both modes. (e) and (f) are respectively simulated and measured performance of a mode-division multiplexing system composed of a dual-mode multiplexer and a demultiplexer based on the ternary pattern. (g) and (h) are respectively SEM images of the fabricated device based on the ternary pattern and the mode-division multiplexing system.

Figures 4(a)–4(d) show the three types of patterns and the simulated insertion loss and crosstalk profiles of ternary pattern, respectively. The simulated insertion loss for both modes is 0.68 dB in average and the crosstalk is less than –25 dB from 1530 nm to 1570 nm. The simulated and measured performance of a mode-division multiplexing system composed of a dual-mode multiplexer and a demultiplexer based on the ternary pattern is illustrated in Figure 4(e) and 4(f), respectively. The simulated insertion loss of this mode-division multiplexing system for both modes is 1.36 dB in average and the crosstalk is less than –20 dB from 1530 nm to 1570 nm. And the measured insertion loss of this mode-division multiplexing system for both modes is 1.51 dB in average and the crosstalk is less than –18 dB over a bandwidth of 40 nm centered at 1550 nm. The SEM pictures of the fabricated dual-mode demultiplexer and the mode-division multiplexing system are given in Figure 4(g) and (h).

The computation times for the designs of 3-dB power divider and dual-mode demultiplexer using the digitized adjoint method are about 1.2 and 7 hours, respectively. For inverse designs of the two same devices using conventional DBS method, the time spent on a single optimization process (convergence of the FOM) are approximate 5.5 and 36 hours, respectively. Meanwhile, the simulated average excess losses over 40 nm wavelength span (1530–1570 nm) of the 3-dB power dividers designed by the digitized adjoint method and DBS method are 0.33 dB and about 0.2 dB, respectively. For the dual-mode demultiplexer designed by the digitized adjoint method, the simulated insertion loss for both modes is 0.68 dB in average and the crosstalk is less than –25 dB from 1530 nm to 1570 nm, while such two parameters for the multiplexer designed by DBS method are 0.47 dB and less than –24 dB from 1530 nm to 1590 nm, respectively. Compared with the brute-force DBS method, the proposed digitized adjoint method could improve the design efficiency by nearly 5 times, and the performance optimization can reach approximately the same level.

In conclusion, the digitized adjoint method is a hybrid of topology optimization and brute-force optimization to improve the efficiency of inverse design of high-performance digital subwavelength nanophotonic devices. Using the proposed method, we have designed and experimentally demonstrated a single-mode 3-dB power divider and a dual-mode demultiplexer with ternary digital patterns based on PhC-like subwavelength structure, respectively. Compared with the DBS brute-force method, the digitized adjoint method increases the design efficiency by nearly 5 times while achieving approximately same device performance. We expect that the digitized adjoint method can be used to design digital nanophotonic devices based on various types of subwavelength structures different from the PhC-like one. By breaking the efficiency bottleneck of the conventional brute-force method with computational time exponentially increasing with the number of pixels and extending the conventional binary pattern to the multi-level pattern, the digitized adjoint method could be applied to inverse design of large-scale digital subwavelength patterns for exploring digital nanophotonic devices with previously unattainable functionality or higher performance.

## ■ METHODS

**Optimization.** A computer with an 8-core CPU (Intel Xeon E5-2637 at 3.5 GHz) and 64 GB memory was employed for simulation. The 3D FDTD simulations were performed via a commercial software (Lumerical FDTD Solutions). We also simulated both types of devices using the inverse design methods described in refs 5 and 11 on the same computer for comparison of design efficiency, respectively. A mesh size of 32.5 nm × 32.5 nm × 30 nm was used in the design of 3-dB power divider. In the design of dual-mode demultiplexer, the mesh size was 30 nm × 30 nm × 30 nm.

**Fabrication.** Both types of devices were fabricated using an electron-beam lithography (EBL) system (Vistec EBPG 5000 Plus) to form the optimum pattern on a SOI platform with a 220 nm thick top silicon layer and 2.0 μm buried oxide layer, and an inductively coupled plasma (ICP) etcher (Plasmalab System 100) to transfer the mask to the silicon device layer based on a single step etching process.

**Characterization.** The transmission profiles were measured by vertical-coupling input and output waveguides. In order to compare with the performance of conventional inverse-designed digital nanophotonic devices, we adopted the same reference structures and experimental setups as refs 5 and 11 to characterize the fabricated 3-dB power divider and dual-mode demultiplexer, respectively.

## ■ AUTHOR INFORMATION

Corresponding Author
*E-mail: mmz@hust.edu.cn.
ORCID
Minming Zhang: 0000-0002-3742-1445
Notes
The authors declare no competing financial interest.

## ■ ACKNOWLEDGMENTS

This work was funded by the National Natural Science Foundation of China (61775069, 61635004) and the Technology Innovation Program of Hubei Province of China (2018AAA037).

## ■ REFERENCES